\documentclass[a4paper,11pt]{article}
\pdfoutput=1

\usepackage{jcappub}

\title{Gravitational waves and neutrino oscillations in Chern-Simons\\ axion gravity}

\author[a,b]{Gaetano Lambiase,}
\author[a,b]{Leonardo Mastrototaro,}
\author[c,d\dag]{\newline and Luca Visinelli\note[\dag]{Corresponding author.}}
\affiliation[a]{Dipartimento di Fisica ``E.R Caianiello'', Universit\`a degli Studi di Salerno,\\ Via Giovanni Paolo II, 132 - 84084 Fisciano (SA), Italy}
\affiliation[b]{Istituto Nazionale di Fisica Nucleare - Gruppo Collegato di Salerno - Sezione di Napoli,\\ Via Giovanni Paolo II, 132 - 84084 Fisciano (SA), Italy}
\affiliation[c]{Astronomy division, Tsung-Dao Lee Institute (TDLI),\\ 520 Shengrong Road, 201210 Shanghai, P.\ R.\ China}
\affiliation[d]{School of Physics and Astronomy, Shanghai Jiao Tong University,\\ 800 Dongchuan Road, 200240 Shanghai, P.\ R.\ China}
\emailAdd{lambiase@sa.infn.it}
\emailAdd{lmastrototaro@unisa.it}
\emailAdd{luca.visinelli@sjtu.edu.cn}

\abstract{We investigate the modifications in the neutrino flavor oscillations under the influence of a stochastic gravitational wave background (SGWB), in a scenario in which General Relativity is modified by an additional Chern-Simons (CS) term. Assuming that the dark matter halo is in the form of axions, the CS coupling modifies the pattern of the neutrino flavor oscillations at Earth up to a total suppression in some frequency range. At the same time, the SGWB in the halo could stimulate the axion decay into gravitons over a narrow frequency range, leading to a potentially detectable resonance peak in the enhanced SGWB strain. A consistent picture would require these features to potentially show up in neutrino detection from supernovae, gravitational wave detectors, and experiments aimed at the search for axions in the Milky Way halo.
}

\begin{document}
\maketitle
\flushbottom

\section{Introduction}
\label{sec:introduction}

Axions are pseudo-scalar particles with an anomalous coupling to gauge bosons~\cite{Georgi:1986df}. When originated from the spontaneous breaking of a global symmetry, axions generally acquire a small mass term that drives the onset of a non-thermal and cold population~\cite{Arvanitaki:2009fg, Arias:2012az, Marsh:2013taa, Visinelli:2017imh, Graham:2018jyp, Takahashi:2018tdu, Ho:2019ayl}. For this reason, these particles are actively sought as an explanation for the dark matter (DM) puzzle over a vast range of masses and couplings with the Standard Model fields~\cite{Preskill:1982cy, Abbott:1982af, Dine:1982ah}. In addition, axions might play a relevant role in the resolution of several problems in particle physics and cosmology~\cite{raffelt} and they could lead to distinctive signatures such as black hole (BH) superradiance, cosmic birefringence, as well as a modified evolution of stellar structures. Ultralight axions of mass $m_\phi\sim10^{-22}\,$eV have been invoked to reconcile some discrepancies between modelling and observed features at the galactic scale~\cite{Hu:2000ke} and can be constructed within fundamental theories such as string theory~\cite{Svrcek:2006yi, Stott:2017hvl, Cicoli:2021gss}.

The experimental search for axions is based on their feeble coupling to Standard Model (SM) particles. The interaction with the electromagnetic field leads to a enriched phenomenology that includes spectral distortions~\cite{Hooper:2007bq, TheFermi-LAT:2016zue, Xia:2018xbt, Liang:2018mqm, Xia:2019yud, Li:2020pcn}, an axion-mediated fifth force~\cite{Jaeckel:2010ni, Rong:2017wzk, Rong:2018yos}, cosmological bounds~\cite{Hlozek:2014lca, Visinelli:2018utg, Dror:2020zru, Caloni:2022uya}, the rotation of the photon polarization angle induced by the axion field~\cite{Carroll:1989vb, Carroll:1991zs, Harari:1992ea, Ivanov:2018byi, Yuan:2021yua, Fujita:2018zaj, Liu:2019brz, Fedderke:2019ajk, Caputo:2019tms, Chen:2019fsq, Chigusa:2019rra, Poddar:2020qft, Basu:2020gsy, Nagano:2019rbw, Choi:2021aze}, and spin-axion coupling~\cite{Flambaum:2009mz}. The interactions with the gluons or with colored quarks would also induce an electric dipole moment of the neutron which can be sensitively probed through designed experimental setups. For recent reviews of experiments see Refs.~\cite{Graham:2015ouw, Irastorza:2018dyq}, while theory reviews are in Refs.~\cite{Marsh:2015xka, DiLuzio:2020wdo}. Despite severe efforts in designing novel experimental setups and astrophysical probes, the elusiveness of the axion coupling to SM particles has proved their detection to be extremely difficult. This motivates the extension of the search to less conventional probes such as gravitational effects, which can nevertheless exploit the flourished development in recent years of the field.

In some contexts, it is predicted that the axion interacts with gravity through a Chern-Simons (CS) term~\cite{Campbell:1990ai, Lue:1998mq, Jackiw:2003pm, Alexander:2004wk, Alexander:2007kv, Alexander:2008wi, Yunes:2009hc, Molina:2010fb}, in a form that resembles the coupling between the axion and SM gauge bosons. See Ref.~\cite{Alexander:2009tp} for a review. Such a coupling is generally expected within a low-energy realization of string or M-theory~\cite{Witten:1984dg, Choi:1985je, Choi:1999zy, Kim:2016ncr, Choi:1999zy, Kim:2016ncr} as a result of anomaly cancellation, and might impact low-energy local gravitational phenomena~\cite{Smith:2007jm}, as well as the propagation of gravitational waves (GW)~\cite{Nojiri:2019nar}. In a cosmological setup, the presence of a CS coupling would not alter the background evolution of the axion, although the gravity CS term would affect the propagation of gravitational waves and induce a modification of primordial perturbations~\cite{Lue:1998mq}. Whenever the GW frequency matches one-half the axion Compton frequency, a GW strain propagating in a coherent axion field would be enhanced through resonant axion decay~\cite{Yoshida:2017cjl, Chu:2020iil, Kitajima:2018zco, Jung:2020aem} and even lead to a burst of GWs from localized axion clumps~\cite{Sun:2020gem}. As a consequence, a resonance peak in the GW spectrum with a frequency related to the axion mass would arise.\footnote{A similar effect also arises for electromagnetic waves propagating through the coherent axion dark matter field~\cite{Yoshida:2017ehj, Rosa:2017ury, Arza:2018dcy, Caputo:2018ljp, Caputo:2018vmy}; see also Refs.~\cite{Arza:2019nta,Arza:2018dcy} for the analysis of echo's, Refs.~\cite{Hertzberg:2018zte, Wang:2020zur} for the destabilization of the axion structures, which may provide signals in the background or produce an explosive burst~\cite{Tkachev:2014dpa}.}

A new framework for investigating these effects explored in recent literature is provided by the stochastic background of GWs (SGWB) or the chirping GW strain produced for example by the merging of compact objects, affecting the probability to detect a given neutrino flavor~\cite{Rosado:2011kv, Giudice:2016zpa, Jung:2017flg, Lai:2018rto, Christian:2018vsi, Dai:2018enj, Jung:2018kde, Choi:2018axi}. In more detail, the energy density in GWs impacts the length scale over which the coherent neutrino oscillations occur so that the detection of the burst of a neutrino flux could test models of the SGWB and chirping GWs. After the recent success in experimental studies, it has been determined that neutrinos are massive particles and have nonzero mixing between different flavors that leads to flavor oscillations. External fields can significantly modify the process of neutrino oscillations and therefore the gravitational interaction can have observable effects on the propagation and oscillation of neutrino flavor~\cite{Ahluwalia:1996ev, Fornengo:1996ef, Cardall:1996cd, Capozziello:1999qm, Lambiase:2004qk, Lambiase:2005gt, Cuesta:2008te, Lambiase:2013haa, Chakraborty:2013ywa, Visinelli:2014xsa, Chakraborty:2015vla, Dvornikov:2019fhi, Dvornikov:2006ji, Buoninfante:2019der, Lambiase:2021txu, Dvornikov:2021hps, Swami:2022xet}. In particular, we will take into account the contribution of GW following Ref.~\cite{Dvornikov:2021hps}. The influence of GWs would open a window for multimessenger astronomy because some dark particle candidates such as the axion may influence their production and evolution.

Neutrino oscillations over astronomical distances transform the flavor composition according to the Pontecorvo–Maki–Nakagawa–Sakata (PMNS) matrix~\cite{Learned:1994wg, Majumdar:2006px, Visinelli:2008ds, Esmaili:2009uq, Choubey:2009jq, IceCube:2015rro}. Using the global best-fit mixing parameters in normal and inverted hierarchy~\cite{Gonzalez-Garcia:2014bfa}, it is possible to exactly determine the neutrino flavor evolution. Therefore, the observation of a ratio inconsistent with the expected flux, provided a precise source production model and a good spatial sensitivity of the detector, would be a signal of new physics in the neutrino sector. For neutrinos with energies $E\sim \mathcal{O}($MeV-PeV), this effect would add to the already studied cases such as neutrino decay~\cite{Beacom:2002vi, Baerwald:2012kc}, sterile neutrinos~\cite{Athar:2000yw}, pseudo-Dirac neutrinos~\cite{Beacom:2003eu, Esmaili:2009fk}, Lorentz or CPT violation~\cite{Hooper:2005jp}, and quantum gravity-induced decoherence~\cite{Anchordoqui:2005gj}.

In this paper, we aim to study the effects of a SGWB on neutrino oscillations, focusing on the effects induced by the GW enhancement at a particular frequency from the resonant decay of the axions in the dark matter halo. Accounting for the GW background provided by known sources such as compact binaries and core-collapse supernovae (SN), we show that a possible suppression of the neutrino coherence length induced by the CS term could be achieved for light axions of mass $\sim (10^{-11} \textrm{--} 10^{-13})\,$eV. This range corresponds to the GW frequency range $\sim (10 \textrm{--} 1000)\,$Hz, making the search a feasible target for interferometers such as the Laser Interferometer Gravitational-Wave Observatory (LIGO) plus VIRGO and the Kamioka Gravitational Wave Detector (KAGRA) network~\cite{Abbott:2016xvh}. At the same time, the suppression in the neutrino oscillation probability would appear in the energy spectrum within $10 \textrm{--} 100\,$MeV and could be probed through the energy spectrum of neutrinos detected by the Jiangmen Underground Neutrino Observatory (JUNO)~\cite{JUNO:2015zny} and by Super-Kamiokande~\cite{Super-Kamiokande:2002weg}.

This paper is organized as follows. In Sec.~\ref{sec:CS} we discuss the theoretical foundations of the CS axion-gravity coupling. In Sec.~\ref{sec:SGWB} we describe the SGWB used in the analysis. In Sec.~\ref{Formulation}, we discuss the formalism for the description of neutrino flavor oscillations and the effects of a GW background. In Sec.~\ref{Results}, we show the enhancement in the GW background and the effect on the coherence length of neutrino oscillations. We consider as a possible astrophysical application the neutrinos emitted from the explosion of a supernova (SN) and detected in Super-Kamiokande or JUNO. The relevant conclusions of this analysis are summarized in Sec.~\ref{Conclusions}. In this paper, we work in units with $\hbar = c = 1$ unless otherwise specified.

\section{Axion-gravity coupling}
\label{sec:CS}

\subsection{Propagation of gravitational waves in Chern-Simons gravity}

We consider the propagation of a pseudoscalar field $\phi$ described by the action
\begin{equation}
    \label{eq:action}
    S = \int {\rm d}^4x\sqrt{-g}\left(\frac{R}{2\kappa^2} - \frac{1}{2}\nabla^\mu \phi \nabla_\mu \phi - \frac{1}{2}m_\phi^2\phi^2 + {\cal L}_{\rm CS}\right)\,,
\end{equation}
where $\kappa \equiv \sqrt{8\pi G}$ and $G$ is Newton's constant, $g^{\mu\nu}$ is the background metric with the determinant $g$, $R$ the Ricci scalar, $R^{\mu\nu}$ the Ricci tensor, and $m_\phi$ the axion mass. Here, ${\cal L}_{\rm CS}$ is the Lagrangian term describing the CS coupling between the axion and gravity arising from the gravitational anomaly,
\begin{equation}
    \label{CSLagrabgian}
    {\cal L}_{\rm CS} = \frac{\ell_{\rm CS}^2}{8\kappa}\, \phi R \widetilde{R} = \frac{\ell_{\rm CS}^2}{8\kappa}\, \phi \varepsilon^{\alpha\beta\gamma\delta} R_{\alpha\beta\rho\sigma}R^{\rho\sigma}_{\quad \gamma\delta}\,,
\end{equation}
with the Levi-Civita tensor $\varepsilon^{\alpha\beta\gamma\delta}$. This CS term is analogous to the Lagrangian for the coupling of the axion with the electromagnetic tensor $F$, ${\cal L} \propto \phi F\tilde F$, with the length scale $\ell_{\rm CS}$ parameterising the effective CS coupling~\cite{Yoshida:2017cjl,Okounkova:2017yby}.

A CS correction to general relativity is expected in models of quantum gravity such as superstring theory and loop quantum gravity. In four-dimensional compactifications of superstring theory, a coupling of the type in Eq.~\eqref{CSLagrabgian} appears almost inevitably due to the Green-Schwarz anomaly cancellation mechanism~\cite{Green:1984sg}, as well as in the presence of Ramond-Ramond scalars due to duality symmetries~\cite{Alexander:2004xd}, with the form~\cite{Lue:1998mq, Choi:1999zy, Alexander:2004us}
\begin{equation}
    {\cal L}_{\rm CS} = \frac{\cal{N}}{2\pi M_S}\, \phi R \widetilde{R}\,,
\end{equation}
where $\phi$ is the axion field arising within the theory, $M_S$ is the string scale and $\cal{N}$ is the number of string degrees of freedom. Setting for example $M_S \sim 1/\kappa$ leads to $\ell_{\rm CS} \sim \ell_{\rm Pl}\,{\cal N}^{1/2}$, where $\ell_{\rm Pl}$ is the reduced Planck length. In loop quantum gravity, once the Barbero-Immirzi parameter is promoted to a pseudoscalar field $\beta$~\cite{Mercuri:2009zi}, the CS coupling appears at the semi-classical level as an interaction between $\beta$ and the gauge bosons with a length scale $\ell_{\rm CS} \sim \kappa$ that is of the order of the Planck's length~\cite{Taveras:2008yf, Mercuri:2009zt}. More generally, the CS term arises in effective theories of gravity once expanding the action in the curvature tensor~\cite{Weinberg:2008hq}.

Ignoring cosmic expansion, the action in Eq.~\eqref{eq:action} considered to second order around a flat Minkowski spacetime $\eta^{\mu\nu}$ as $g^{\mu\nu} = \eta^{\mu\nu} + h^{\mu\nu}$ with $|h^{\mu\nu}|\ll 1$ leads to
\begin{eqnarray}
    \label{eq:EHaction}
    S &\supset& \frac{1}{8\kappa^2} \int d^4 x \bigg[\dot h^i{}_j \dot h^j{}_i- (\partial_k h^i{}_j)(\partial^k h^j{}_i) \label{actionCS} \\
    & &  - \kappa\ell_{\rm CS}^2 \dot\phi \epsilon^{ijk}\left(\dot h^l{}_i\partial_j\dot h_{kl}-(\partial^mh^l{}_i) (\partial_m\partial_jh_{kl})\right) \bigg]\,, \nonumber
\end{eqnarray}
where a dot is a derivation with respect to time and $h_{ij}$ is the space-space component of the tensor of metric perturbations, following the equations of motion derived from the linearized action above as~\cite{Alexander:2004wk, Alexander:2009tp}
\begin{equation}
    \label{waveeq}
    \left(\partial_t^2-\nabla^2\right) h^j_i  = \kappa\ell_{\rm CS}^2 \epsilon^{lkj} \left(\ddot\phi \partial_l\dot{h}_{ki}+\dot\phi \partial_l\ddot{h}_{ki} -\dot\phi \nabla^2\partial_l h_{ki}\right)\,.
\end{equation}

Overall, the cosmological model is not affected by the presence of the CS term since the flat Friedmann-Robertson-Walker metric is a solution to the CS gravity. At the same time, the equations for the axion field are not modified by the presence of the CS term up to the second order in perturbations~\cite{Haghani:2017yjk}, and the CS term enters the modifications of metric perturbations~\cite{Lue:1998mq, Alexander:2004us}. Here, we do not discuss the effects of the CS term in the early Universe further, and we assume that the axion is the DM regardless of the specific details of the production mechanisms.

In more detail, we assume that the dark matter halo in the Milky Way (MW) is composed of cold axions that coherently oscillate with a frequency equal to their rest mass,
\begin{equation}
    \phi(t)  =  \sqrt{\frac{\rho_{\rm DM}}{2m_\phi^2}} e^{-i m_\phi t} + \text{c.c.}\,,
\end{equation}
where the amplitude of the oscillations is related to the local energy density of dark matter in the halo $\rho_{\rm DM}$.

DM axions are a highly coherent superposition of axion waves and are characterized by a long spatial coherence over a patch of length scale $L_{\rm patch} \sim 1/ (m_\phi \Delta v)$, with a small velocity dispersion $\Delta v\sim 10^{-3}$ related with the non-relativistic nature of DM~\cite{Seidel:1990jh, Kolb:1993zz, Sikivie:2009qn, Guth:2014hsa}. In the case in which the GW frequency matches the axion Compton frequency, the stochastic (chirping) GW strain can be enhanced at the resonance by the coherent (temporal) oscillation, leading to a burst of gravitons at the frequency $f_{\rm GW} \equiv m_\phi/(4\pi)$ and a corresponding energy $\omega_{\rm GW} = 2\pi f_{\rm GW} = m_\phi/2$. Such a resonance peak could be discernible from the waveform of the chirping GW and noise because all observed GWs with the same frequency would undergo a similar phenomenon. Ref.~\cite{Jung:2020aem} considers the effects of a CS coupling in relation to the chirping GWs detectable in the LIGO-VIRGO band, for which the temporal coherence $\Delta t_{\rm coh}$ of the axions within a coherent patch is much larger than the duration $\Delta t_{\rm GW}$ of the chirping GW pulse. Indeed, the correlation of all 11 GW observations at LIGO/VIRGO O1+O2 allows to strongly constrain the CS coupling~\cite{Jung:2020aem}.

The solutions of Eq.~\eqref{waveeq} for a strain of GWs propagating through an axion DM halo read~\cite{Jung:2020aem}
\begin{eqnarray}
    h_A^{(s)}(t) &=& h_0^{(s)}\left(\cosh(\mu t) - i \epsilon \frac{m_\phi}{2\mu} \sinh(\mu t) \right), \label{solF} \\
    h_B^{(s)}(t) &=& i {h_0^{(s)}}^{*}\,\lambda^{(s)} \gamma e^{i\Phi}\,\frac{m_\phi}{2\mu}\sinh(\mu t)\,, \label{solB}
\end{eqnarray}
where the suffixes $A$, $B$ stand for forward and backward waves respectively, $s$=R ($s$=L) indicates the right-handed (left-handed) helicity of the GW with eigenvalues $\lambda^{(\rm R)}=+1$ $(\lambda^{(\rm L)}=-1)$ and $\Phi$ is the phase of the axion field. The initial conditions to Eqs.~\eqref{solF}--\eqref{solB} correspond to a forward-propagating chirping GW with $h_B^{(s)}(0)=0$ and $h_A^{(s)}(0) = h_0^{(s)}$, with the initial amplitude $h_0^{(s)} \equiv h^{(s)}(t=0)$. We have defined the enhancement rate $\mu$, the fractional deviation $\epsilon$ of the signal $f$ from the resonance frequency $f_{\rm GW}$, and the dimensionless parameter $\gamma$, as
\begin{eqnarray}
    \mu &\equiv& \frac{m_\phi}{2}\,\sqrt{\gamma^2 - \epsilon^2}\,,\\
    \epsilon &\equiv& \frac{f - f_{\rm GW}}{f_{\rm GW}}\,,\\
    \gamma & \equiv & \sqrt{\pi G \rho_{\rm DM}}\,\ell_{\rm CS}^2 m_\phi\,.\label{gamma}
\end{eqnarray}

For finite propagation of GW with $\mu t \ll 1$, where $t$ is the propagation time, Eqs.~\eqref{solF}--\eqref{solB} read
\begin{equation}
    \label{eqhF}
    h_A^{(s)}(t)  = h_0^{(s)} \, e^{-i \psi(t)}\,\left(1 + \delta(t)\right)\,,
\end{equation}
where the amplitude enhancement and the phase of the wave within the individual patch in which the axion field is coherent are expressed respectively as
\begin{eqnarray}
    \delta(t) &\approx& \frac{\gamma^2}{2} \left(\frac{m_\phi}{2} t\right)^2\!\text{sinc}^2\!\left(\frac{m_\phi}{2} \epsilon  t \right)\,,\label{eqforF}\\
    \psi(t) &\approx& \frac{m_\phi}{2}\epsilon t \left\{1 \!+\! \frac{1}{2}\left[\text{sinc}\left(m_\phi\epsilon t\right) -1 \right] \left(\frac{\gamma}{\epsilon}\right)^2\! \right\},\label{psi}
\end{eqnarray}
where we introduced the function $\text{sinc}(x) = \sin(x)/x$. Inserting the time spent by the GW within the coherent patch of size $L_{\rm patch} \sim 1/(m_\phi \Delta v)$ leads to the amplitude enhancement in each patch and in the frequency domain as
\begin{equation}
    \label{eqforFpatch}
    \delta(f) = \frac{\gamma^2}{8\Delta v^2} \text{sinc}^2 \left(\frac{\epsilon}{2 \Delta v} \right)\,.
\end{equation}
The frequency width of the enhancement occurs around the central peak of the sinc function, $-2 \Delta v \lesssim \epsilon \lesssim 2 \Delta v$, corresponding to the frequency range in GWs,
\begin{equation}
    m_\phi/2 - m_\phi \Delta v \lesssim \omega_{\rm GW} \lesssim m_\phi/2 + m_\phi \Delta v\,,
\end{equation}
leading to the peak width $\sim 2 m_\phi \Delta v$. In the context of this work, the expansion of the expression in Eq.~\eqref{eqhF} is justified since $\mu L_{\rm patch} \ll 1$ for the axion mass $m_\phi \lesssim 10^{-8}\,$eV, which is well within the range of axion masses $m_\phi = (10^{-13} - 10^{-11})\,$eV that we find to be relevant for GW detection.

We comment on the quenching of the enhancement and the energy conservation in the system. The energy carried by the forward scattered waves in Eq.~\eqref{solF} is
\begin{equation}
    \Delta \rho_{\rm GW} = \frac{\omega_{\rm GW}^2}{4\kappa^2} \sum_s \left(|h_A^{(s)}(t)|^2 - |h_A^{(s)}(0)|^2\right) = \frac{\omega_{\rm GW}^2}{4\kappa^2} \,\frac{\gamma^2}{4\Delta v^2} \,h_0^2\, \text{sinc}^2 \left(\frac{\epsilon}{2 \Delta v} \right)\,,
\end{equation}
where in the last expression we used Eq.~\eqref{eqhF} with the approximation $\delta \ll 1$ and $h_0^2 = \sum_s |h_0^{(s)}|^2$. Generally, the GW enhancement in a single patch is a negligible contribution to the stochastic GW background discussed in Sec.~\ref{sec:SGWB} since $\Delta \rho_{\rm GW}/\rho_{\rm GW} < \delta \ll 1$. Following the discussion in Ref.~\cite{Jung:2020aem}, the energy loss in axions with a patch amounts to $|\Delta \rho_\phi| = 2\Delta \rho_{\rm GW}$, so that the backreaction on the axion density can be quantified in terms of percentage loss
\begin{equation}
    \frac{|\Delta \rho_\phi|}{\rho_\phi} = \frac{2\Delta \rho_{\rm GW}}{\rho_{\rm DM}} \leq \frac{\omega_{\rm GW}^4 \ell_{\rm CS}^4}{32\Delta v^2}\,h_0^2 = \frac{3 H_0^2 m_\phi^2 \ell_{\rm CS}^4}{64\Delta v^2}\,\Omega_{{\rm GW},0}\,,
\end{equation}
where the mean energy density is obtained from Eq.~\eqref{gamma} and we used the fact that $|\text{sinc}(x)| \leq 1$. The last expression relates the strain $h_0$ to the GW fractional density in the absence of the enhancement $\Omega_{{\rm GW},0}$ in terms of the Hubble constant $H_0$. Setting $\ell_{\rm CS} = 10^8\,$km and $m_\phi = 10^{-12}\,$eV, the bound reads $|\Delta \rho_\phi|/\rho_\phi \lesssim \Omega_{{\rm GW},0}/10^{13}$.

The existence of a gravitational CS coupling would lead to various astrophysical phenomena which can be used to bound the parameter $\ell_{\rm CS}$. Tests of frame-dragging effects around the Earth by Gravity Probe B and the LAGEOS satellites bound the quantity~\cite{Smith:2007jm}
\begin{equation}
    \label{eq:smith}
    m_{\rm CS} \equiv \frac{2}{\kappa \ell_{\rm CS}^2 \dot\theta} \lesssim 2\times 10^{-13}{\rm\,eV}\,.
\end{equation}
Once set $\dot\theta = \sqrt{\rho_{\rm DM}/2}$, Eq.~\eqref{eq:smith} translates into an upper bound on the CS length scale as $\ell_{\rm CS} \lesssim 2\times 10^{-2}\,$pc.
A more stringent bound is obtained when considering the modification induced by the CS coupling to the rate of periastron precession in a system of double binary pulsars~\cite{Ali-Haimoud:2011wpu}, see also Refs.~\cite{Yunes:2009ch}, which leads to
\begin{equation}
    \label{eq:yunes}
    m_{\rm CS} \lesssim 5\times 10^{-10}{\rm\,eV}\,,
\end{equation}
which translates into the bound $\ell_{\rm CS} \lesssim 5\times 10^{-4}\,$pc. Finally, revisiting the bounds from frame-dragging effects leads to $\xi^{1/4}\lesssim 10^8\,$km~\cite{AliHaimoud:2011fw, AliHaimoud:2011fw}, where $\xi \equiv \ell_{\rm CS}^4/2$, thus leading to
\begin{equation}
    \label{eq:alihaimoud}
    \ell_{\rm CS} \lesssim 1.2\times 10^{8}{\rm\,km}\,.
\end{equation}
Future measurements with an accuracy below 10\% of the moment of inertia for two pulsars in a binary system could improve the bound to $\ell_{\rm CS} \lesssim 10^5\,$km using the forecast sensitivity for LISA or even to $\ell_{\rm CS} \lesssim 10\,$km using the forecast sensitivity for DECIGO~\cite{Yagi:2012vf}.

\subsection{GW enhancement in the Galaxy}

We consider an event that occurs on the light-of-sight $s$ from Earth, so that the distance with respect to the Galactic Center (GC) depending on the galactic coordinates $(b, l)$ is $r(s) = (s^2+r_\odot^2-2sr_\odot \cos b\cos l)^{1/2}$. For a DM halo extending to the virial radius $R_{\rm vir}$, the total distance $D$ travelled inside the halo to Earth is obtained by inverting $r(D) = R_{\rm vir}$. Since all of the axion signals from one patch contribute to the stimulation in the next patch, a cumulative enhancement occurs after crossing a number $N$ of patches. The cumulative effect $A_{\rm enh} \equiv \Omega_{\rm GW}/\Omega_{\rm GW,0}$ of the enhancement from all coherent patches in the DM halo, each contributing as in Eq.~\eqref{eqforFpatch}, along the line of sight $s$ is then
\begin{eqnarray}
    \label{eq:totenhancement}
    A_{\rm enh} &=& \prod_{i=1}^{N} \left(1+\delta_i(f)\right) = \exp\left(\sum_{i=1}^{N}\log (1+\delta_i(f))\right)\\
    &\approx& \exp\left(\sum_{i=1}^{N} \delta_i (f)\right) \approx \exp\left(\int_0^D {\rm d}s\,\frac{\delta(f)}{L_{\rm patch}(r(s))}\right)\,,\nonumber
\end{eqnarray}
where the index $i$ labels the $i$-th patch at a distance $r_i$ from the GC and $L_{\rm patch}(r) = [m_\phi \Delta v(r)]^{-1}$ is the coherence length. In the last step, we have assumed that the discrete distribution of the patches in the Galaxy can be approximated by a continuous function. For this, $\delta(f)$ is a function of the radial distance $r$ from the GC through the velocity dispersion $\Delta v$ and the DM density $\rho_{\rm DM}$. Inserting Eqs.~\eqref{gamma}--\eqref{eqforFpatch} into Eq.~\eqref{eq:totenhancement} we finally obtain
\begin{equation}
    \label{ndelta1}
    A_{\rm enh} \!=\! \exp\left[\int_0^D {\rm d}s\frac{\pi G \rho_{\rm DM}(r(s))}{8\Delta v}\ell_{\rm CS}^4 m_\phi^3{\rm sinc}^2\left(\frac{\epsilon}{2\Delta v}\right)\right]\,. 
\end{equation}
This expression will play a central role in the analysis of the neutrino oscillations interacting with GW coupled to axion in the MW. Note, that the expression in Eq.~\eqref{ndelta1} represents an improvement over the results in Ref.~\cite{Jung:2020aem} since we have assumed that the dispersion velocity is radial-dependent and the observer is not placed at the center of the DM distribution. This latter modification allows us to include the directionality of the signal in the study.

In the analysis, we use the expression in Eq.~\eqref{ndelta1} to derive the main results. However, it is useful to consider an approximate expression for the enhancement to discuss the results. Since the width of the distribution in Eq.~\eqref{psi} is given by the momentum dispersion $2m_\phi\Delta v$, the expression for the enhancement in Eq.~\eqref{ndelta1} with the patch distribution $\delta_{\rm patch}(f)$ can be approximated near the resonance as
\begin{eqnarray}
    \label{eq:enh_approx}
    A_{\rm enh} &=& 1 \!+\! \left(e^{\ell_{\rm CS}^4 m_\phi^3\alpha_1}\!-\!1\right)\!e^{-\ell_{\rm CS}^4 m_\phi^3\alpha_2 \frac{(f \!-\! f_{\rm GW})^2}{2 f_{\rm GW}^2}},\\
    \alpha_1 &=& \int_0^D{\rm d}s\frac{\pi G \rho_{\rm DM}}{8 \Delta v}\,,\label{def:alpha1}\\
    \alpha_2 &=& \int_0^D{\rm d}s\frac{\pi G \rho_{\rm DM}}{64 \Delta v^3}\,,\label{def:alpha2}
\end{eqnarray}
where $f_{\rm GW} = m_\phi/(4\pi)$ is the frequency of each emitted graviton. Away from the resonance, there is no enhancement and $A_{\rm enh} \approx 1$, while around the peak the function is approximated by a delta distribution.

In the following, we model the DM distribution in the MW as a spherical distribution with a Navarro-Frenk-White radial profile~\cite{Navarro:1995iw},
\begin{equation}
    \label{eq:nfw}
    \rho_{\rm DM}(r) = \rho_s\,\frac{r_s}{r}\,\left(1+\frac{r}{r_s}\right)^{-2}\,,
\end{equation}
where $r_s\approx 16\,$kpc~\cite{Nesti:2013uwa} and we fix the normalization by requiring that the density at the Solar system position, $r_\odot \approx 8.5\,$kpc, is $\rho_\odot = 0.4{\rm\,GeV\,cm^{-3}}$, leading to $\rho_s \approx 0.5{\rm\,GeV\,cm^{-3}}$. Assuming that DM is on circular orbits, a parcel at distance $r$ from the GC has velocity dispersion
\begin{equation}
    \Delta v(r) = \sqrt{\frac{4\pi G \rho_s r_s^3}{r}}\left[\ln(1+r/r_s) - \frac{r/r_s}{1+r/r_s}\right]^{1/2}\,,
\end{equation}
and we fix the concentration parameter for the MW as $C = 10$, so that the virial radius is $r_{\rm vir} \equiv Cr_s = 160\,$kpc and the total DM mass obtained from integrating Eq.~\eqref{eq:nfw} up to $r_{\rm vir}$ is $M_{\rm DM}\approx 1\times 10^{12}\,M_\odot$.

\section{Stochastic GW background}
\label{sec:SGWB}

The collective GW signals from all incoherent sources in the Universe contribute to a SGWB at various frequency windows. Understanding the shape of the SGWB is of fundamental importance to test various cosmological theories as well as the physics and astrophysics of compact objects. Such a background is a primary target for the next generation of GW detectors including interferometers such as the advanced LIGO (aLIGO)/VIRGO and LISA, as well as the search at much lower frequencies from pulsar timing array. Indeed, we already have overwhelming confirmation of the coalescence of binary systems of neutron stars (NSs) and BHs of stellar origin, which should lead to a SGWB whose properties would reveal much about the distribution and history of these populations of objects. At the same time, the Event Horizon Telescope has confirmed the existence of supermassive BHs (SMBHs) from the imaging of the dark shadow surrounding the compact objects at the core of M87$^*$ and Sgr A$^*$. At much lower frequencies accessible with pulsar timing array techniques, we expect a SGWB from the merging of SMBHs in the Universe. Exotic components are also expected to lead to a SGWB that is potentially of similar amplitude, including cosmic strings, phase transitions in the early Universe, or inflation. For a list of possible sources see Refs.~\cite{Schneider:2010ks, Kuroyanagi:2018csn}.

Here, we focus on the effects of known compact objects to the SGWB amplitude, for which the root mean square of the GW strain averaged over the polarization and the phase is~\cite{Rosado:2011kv, Christensen:2018iqi}
\begin{equation}
    \label{h2estimate}
    \langle h^2\rangle = \int {\rm d}f \frac{3}{\pi^2}\,H_0^2\,\frac{\Omega_{\rm GW}(f)}{f^2}\,,
\end{equation}
where the appearance of the Hubble constant $H_0 \equiv 100h\,$km/s/Mpc in the prefactor of Eq.~\eqref{h2estimate} leads to the GW interferometry experiments being sensitive on the combination of the GW density parameter multiplied by $h^2$, or $\Omega_{\rm GW}h^2$. The strain is generally much larger than the amplitude $h_{\mathrm{eff}}=10^{-22}$ associated with the chirping GW from a single binary merger.

We first discuss the SGWB expected from various populations of binary systems. The spectrum of GWs radiated from a binary system of two compact objects with total mass $M_{\rm tot}$ and compactness $\mathcal{C}$ during the inspiral phase is given by $\Omega_{\rm GW}(f) \propto f^{2/3}$, peaking at the frequency $f_{\rm ISCO}$ corresponding to the innermost stable circular orbit (ISCO) as $f_{\rm ISCO} \approx (1/\pi) \mathcal{C}^{3/2}/M_{\rm tot}$~\cite{Farmer:2003pa, Meacher:2015iua}. For two NSs each of mass $M_{\rm NS} = 1.4\,M_\odot$, this corresponds to $f_{\rm ISCO} \approx 1600\,$Hz with a strain corresponding to $\Omega_{\rm GW} \sim 10^{-9}$, with binary system of heavier counterparts having smaller frequencies. Binary NS system dominate the SGWB for frequencies $f \gtrsim 0.1\,$Hz.

A possible, even larger contribution to the SGWB at the kHz frequency range could come from binary systems of magnetars, rapidly-rotating NSs surrounded by a large magnetic field which could be responsible for a large GW emission when the rotation axis is not aligned with the magnetic axis. More in depth, a magnetar possesses both an internal magnetic field with toroidal and poloidal components, and an external magnetic field that is predominantly poloidal. The intense inner magnetic field induces a significant quadrupolar deformation which would alter the strain of GW components when these compact objects are in binary systems. The most promising contribution to the SGWB from binary magnetars in the Universe comes in models where the internal magnetic field is mostly toroidal with strength $B_p \sim 10^{14}\,$G~\cite{Marassi:2010wj}.

Astrophysical compact objects of exotic nature could also form binary systems and further enhance the SGWB strain and dominate the spectrum at various frequency ranges. One such example is BHs formed primordially (see Refs.~\cite{Carr:2020xqk, Green:2020jor} for recent reviews), for which the authors in Ref.~\cite{Bavera:2021wmw}  find that the SGWB would not be altered significantly. However, the use of a different mass spectrum resulting from formation mechanisms other than what is considered in Ref.~\cite{Bavera:2021wmw} could change the outcome and possibly lead to a SGWB at even higher frequencies than those considered here. Compact objects such as boson stars (see Refs.~\cite{Palenzuela:2017kcg, Visinelli:2021uve}) could also contribute to the SGWB~\cite{Croon:2018ybs}, to the extent that the inclusion of new template banks along with those already used for the search of BBH and BNS events by the LIGO and VIRGO detectors could realistically lead to the discovery of these exotic components~\cite{Chia:2020psj, Coogan:2022qxs}. New light bosons could also show up in GW detectors through the superradiant instability induced in the evolution of the boson cloud around rotating BHs (see e.g.\ Refs.~\cite{Yuan:2021ebu}). Here, we do not include these exotic sources in the analysis and we concentrate on the conservative binary systems explored in the previous paragraphs.

A different source of GWs of high frequency results from the end of the life cycle of massive stars of mass above $8M_\odot$, which exhaust their fuel and create an iron core at the inner shell, see e.g.\ Refs.~\cite{Janka:2012wk, Mezzacappa:2020oyq}. Gravity drives the core-collapse once it exceeds the Chandrasekhar limit $\sim 1.4 M_\odot$, leading to a copious release of neutrinos from the core collapse~\cite{Janka:2017vcp} and to the formation of a protoneutron star whose oscillations against gravity, surface, or pressure release a strain of high-frequency GWs. The contribution from all core-collapse supernovae of astrophysical and cosmological origin contribute to a SGWB~\cite{Finkel:2021zgf}.

\section{Neutrino oscillations in the presence of a SGWB}
\label{Formulation}

We now consider the implications of the previous discussions on the propagation of neutrinos, focusing on the effects of a SGWB on neutrino flavor oscillations.\footnote{In principle, the CS term would induce an effective neutrino-axion coupling mediated through gravity~\cite{Alexander:2008wi}. However, for the propagation of neutrinos in the MW these effects can be largely neglected as shown in Appendix~\ref{appendix}. Self-interacting neutrino pairs might act as the CS pseudoscalar and form condensates that interact with propagating neutrinos~\cite{Alexander:2022cow}.} We consider a system of three neutrino generations $\nu_\alpha$ with flavors $\alpha = e, \mu, \tau$, related to the mass eigenstates $\nu_i$ with $i =1,2,3$ by a unitary neutrino mixing matrix $U$ as $\nu_\alpha = U_{\alpha i}\nu_i$. We follow the usual parametrization of the neutrino mixing matrix for Dirac neutrinos,
\begin{equation}
    U=
    R_x(\theta_{23})
    \left(
    \begin{matrix}
    c_{13} &0 &s_{13}e^{-i\delta_{\rm CP}}\\
    0 &1 &0\\
    -s_{13}e^{i\delta_{\rm CP}} &0 &c_{13}\end{matrix}\right)
    R_z(\theta_{12})\,,
    \label{eq:mixingmatrix}
\end{equation}
where $c_{ij}=\cos\theta_{ij}$, $s_{ij}=\sin\theta_{ij}$, $\theta_{ij}$ are the corresponding vacuum mixing angles between flavors $i$ and $j$, $\delta_{\rm CP}$ is the CP violating phase, and the matrices $R_x(\theta)$ and $R_z(\theta)$ describe a rotation by an angle $\theta$ around the axis $x$ and $z$, respectively.

Since the GW strain interacting with the neutrinos has a random phase, orientation and amplitude, it is convenient to switch to the description in terms of the neutrino density matrix $\rho$, whose evolution in terms of the effective Hamiltonian in the vacuum in the mass base $H_0$ and the interaction term $H_{\rm int}$ is $i\dot\rho = [H_0+H_{\rm int}, \rho]$. More precisely, the form of these Hamiltonian operators is $H_0 = UH^{({\rm vac})}U^\dag$ and $H_{\rm int} = UH^{(g)}U^\dag$, with
\begin{eqnarray}
    H^{({\rm vac})} &=& \frac{1}{2E}\mathrm{diag}\left(0,\Delta m_{12}^2,\Delta m_{13}^2\right) \,,    \label{eq:mass-matrix}\\
    H^{(g)} &=& H^{({\rm vac})} \left(A_+ h_+ + A_\times h_\times\right)\,,\label{eq:interaction}
\end{eqnarray}
and with the mass squared difference $\Delta m_{ij}^2=m_i^2-m_j^2$ ($i,j = 1,2,3$). Switching to the interaction picture with the density matrix $\rho_{\rm int} \equiv e^{iH_0t}\rho e^{-iH_0t}$ and taking the average over the SGWB parameters denoted by square brackets leads to the equation of evolution for the neutrino density matrix $\langle\rho\rangle$~\cite{Loreti:1994ry, Dvornikov:2019jkc, Dvornikov:2020dst, Dvornikov:2021sac},
\begin{equation}
     \label{evolution}
     \frac{\rm d}{{\rm d}t}\langle\rho_{\rm int}\rangle (t) = -g(t)\,[H_0,[H_0,\langle\rho_{\rm int}\rangle(t)]] \, ,
\end{equation}
where the function $g(t)$ depends on the strain $h_A(t)$ with polarization $A \in \{+, \times\}$~\cite{Dvornikov:2021sac},
\begin{equation}
    g(t) = \frac{3}{128} \sum_A\int_0^t {\rm d}t' \,\langle h_A(t)h_A(t')\rangle\,.
\end{equation}

To simplify the discussion and the numerical coding for the neutrino propagation, we write Eq.~\eqref{eq:mass-matrix} as $H^{({\rm vac})} \equiv M(\zeta)/L_{\rm osc}$, where we introduce the ratio $\zeta \equiv \Delta m_{12}^2/\Delta m_{13}^2$ and the notation
\begin{equation}
    \label{eq:mass-matrix1}
    M(\zeta) = \mathrm{diag}\left(0,\zeta,1\right)\ ,
\end{equation}
so that the mass matrix in Eq.~\eqref{eq:mass-matrix} can be written in terms of the oscillation length $L_{\rm osc} = 2E/\Delta m_{13}^2$ over which the neutrino mass eigenstates $1\to 3$ convert. We set the neutrino mass square differences $|\Delta m^2_{13}| = 2.449\times 10^{-3}{\rm\,eV^2}$ and $\Delta m^2_{12} = 7.39\times 10^{-5}{\rm\,eV^2}$~\cite{Zyla:2020zbs}, so that $\zeta \approx 0.03$ and the oscillation length is $L_{\rm osc} \sim 2.4\,$km for the typical neutrino energy emitted during a supernova event $\bar E_\nu \sim 15\,$MeV~\cite{Beacom:2010kk}. For comparison, the oscillation length for the mass square difference $\Delta m^2_{12}$ and the same neutrino energy is $L_{\rm osc} \sim 80\,$km.

For stochastic GWs, the root-mean-square of the GW strain is related to the fractional density in GWs at frequency $f$ as in Eq.~\eqref{h2estimate},
\begin{equation}
    \sum_A\langle h_A(t)h_A(t')\rangle = \frac{3 H_0^2}{\pi^2}\int {\rm d}f\,\frac{\Omega_{\rm GW}(f)}{f^3}\,,
\end{equation}
so that Eq.~\eqref{evolution} can be cast as
\begin{equation}
     \label{evolution1}
     \frac{\rm d}{{\rm d}t}\langle\rho_{\rm int}\rangle = -\Gamma_{\rm GW}\,[M(\zeta),[M(\zeta),\langle\rho_{\rm int}\rangle]]\,.
\end{equation}
In the last expression, we used the definition for the dimensionless mass matrix in Eq.~\eqref{eq:mass-matrix1} and we have introduced the rate
\begin{equation}
    \label{eqevolutionrho}
    \Gamma_{\rm GW} = \frac{9}{256\pi^3}\frac{H_0^2}{L_{\rm osc}^2}\int {\rm d}f\,\frac{\Omega_{\rm GW}(f)}{f^4}\,.
\end{equation}
The appearance of the rate $\Gamma_{\rm GW}$ allows us to introduce the neutrino coherence length $L_{\rm coh} \equiv 1/\Gamma_{\rm GW}$ over which the conversion between different flavors occurs. As the GW spike produced by the CS coupling could enhance the integral in Eq.~\eqref{eqevolutionrho} significantly, the coherence length could be reduced in the presence of such an effect. We quantify these results in the following Section.

\section{Results}
\label{Results}

\subsection{Effects on GW propagation}

We now turn to the computation of the flavor mixing rate in Eq.~\eqref{eqevolutionrho}. We rely on the SGWB generated by the merging of known compact objects such as BHs, NSs, and white dwarfs (WDs). Here, we consider the analysis in Ref.~\cite{Rosado:2011kv} that takes into account the SGWB generated by these compact objects, spanning a wide range of frequencies $f \supset [10^{-11} \textrm{--} 10^3 ]\,$Hz (see also Refs.~\cite{Marassi:2011si, Zhu:2012xw, Dvorkin:2016okx}). The lower end of the range is dominated by coalescent binaries of SMBHs with masses in the range $[10^2 - 10^{10}]\,M_\odot$, up to frequencies $\sim 10^{-2}\,$Hz above which compact NS-NS and BH-NS binaries lead to the dominant SGWB contribution; binary WD-WD systems dominate in the narrow window [0.01-0.1]\,Hz. We neglect the effect due to exotic components of astrophysical origin such as boson stars~\cite{Palenzuela:2017kcg, Visinelli:2021uve} or cosmological origin such as a network of cosmic strings~\cite{Vilenkin:1981kz, Hindmarsh:2011qj, Cui:2018rwi, Ramberg:2019dgi}.

The length of propagation over which the neutrino packet is coherent $L_{\rm coh}$ has to be compared with the same quantity in the absence of a CS coupling, $L_{\rm coh,0} = 1/\Gamma_{\rm GW,0}$. The presence of the CS coupling acts as an enhancement of the SGWB at a particular frequency that leads to a smaller value of $L_{\rm coh}$. The effects of the GW can be assessed by the quantity
\begin{equation}
    \label{def_deltaL}
    \delta_L \equiv \frac{L_{\rm coh,0} - L_{\rm coh}}{L_{\rm coh,0}} = 1 - \frac{\Gamma_{\rm GW,0}}{\Gamma_{\rm GW}}\,,
\end{equation}
which is $0 \leq \delta_L \leq 1$, with $\delta_L$ = 0 corresponding to the absence of a CS coupling effect.

\begin{figure}
	\centering
	\includegraphics[width=0.7\textwidth]{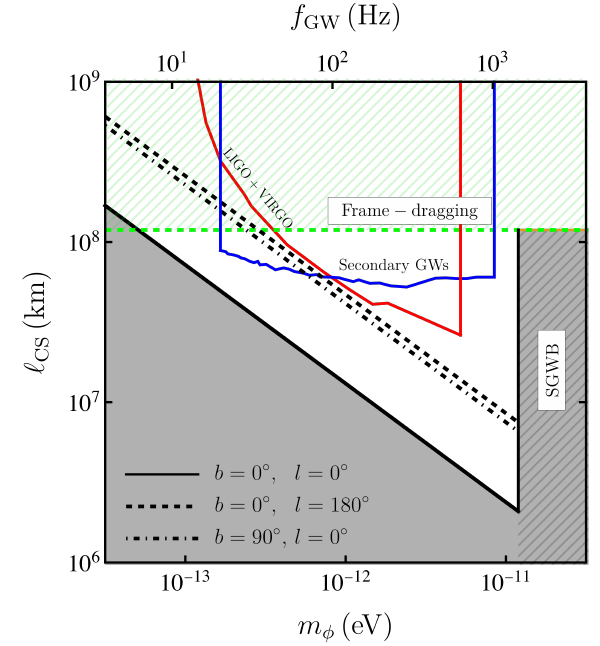}
	\caption{The quantity $\delta_L$ defined in Eq.~\eqref{def_deltaL} as a function of the CS coupling $\ell_{\rm CS}$ (vertical axis) and the axion mass $m_\phi$ (horizontal axis), for different directions of the signal: GC (solid line), Galactic anticenter (dashed line), and perpendicular to the Galactic plane (dot-dashed line). White: $\delta_L = 1$. Gray: $\delta_L = 0$. Also shown (green hashed region) are bounds from the reduction of frame-dragging effects due to the CS coupling~\cite{AliHaimoud:2011fw, AliHaimoud:2011fw}, from the bound obtained from the resonant enhancement of the GW from binary systems~\cite{Jung:2020aem}, and from the search for secondary GWs associated with the coalescence of compact object binaries~\cite{Tsutsui:2022zos}.}
	\label{fig:figurebound}
\end{figure}

Fig.~\ref{fig:figurebound} shows the value of $\delta_L$ as a function of the CS coupling $\ell_{\rm CS}$ (vertical axis) and the axion mass $m_\phi$ (horizontal axis), for different directions of the signal coming from the GC (solid line), Galactic anticenter (dashed line), and perpendicular to the Galactic plane (dot-dashed line). We fix the ratio of the mass difference squared $\zeta = 0.0327$ in Eq.~\eqref{eq:mass-matrix1}. In the grey shaded region outside of the wedge, $\Gamma_{\rm GW} = \Gamma_{\rm GW,0}$ and the value of $\delta_L$ is zero, while inside of the wedge we obtain $\Gamma_{\rm GW} \gg \Gamma_{\rm GW,0}$ and the value of $\delta_L$ is one, with a sharp transition between these regions. The reason can be understood as follows: although the derivation of our results is based on the integration in Eq.~\eqref{eqevolutionrho} over the whole range of frequencies considered, some properties can be predicted analytically.
In particular, in the absence of the enhancement, the integrand is dominated by the lowest frequency in the system $f_{\rm min} \sim 10^{-10}\,$Hz, giving the decay rate
\begin{equation}
    \label{eq:decayrate0}
    \Gamma_{\rm GW, 0} \approx \frac{3}{256 \pi^3} \frac{H_0^2}{L_{\rm osc}^2} \frac{\Omega_{\rm GW}(f_{\rm min})}{f_{\rm min}^3}\,.
\end{equation}
For $f_{\rm min} = 10^{-10}\,$Hz and $\Omega_{\rm GW}(f_{\rm min}) \approx 7\times 10^{-12}$, the neutrino coherence length is $L_{\rm coh,0} \equiv 1/\Gamma_{\rm GW, 0} \sim 7.7{\rm\,pc}\,(L_{\rm osc}/{\rm km})^2 \sim 44\,$pc, where in the last step we have set $L_{\rm osc} = 2.4\,$km. Once the enhancement is considered, the approximation in Eq.~\eqref{eq:enh_approx} leads to the expression
\begin{equation}
    \Gamma_{\rm GW} \approx \frac{9\sqrt{2\pi\alpha_2}}{256 \pi^3} \frac{H_0^2}{L_{\rm osc}^2} \ell_{\rm CS}^2 m_\phi^{3/2} e^{\ell_{\rm CS}^4 m_\phi^3\alpha_1}\frac{\Omega_{\rm GW}(f_{\rm GW})}{f_{\rm GW}^3}\,,
    \label{decayrateCS}
\end{equation}
so that the condition $\Gamma_{\rm GW} > \Gamma_{\rm GW, 0}$ is satisfied when
\begin{equation}
    \label{eq:approx}
    \ell_{\rm CS}^4 m_\phi^3 = \frac{1}{\alpha_1}\ln\left(\frac{1}{\sqrt{18 \pi \alpha_2 \ell_{\rm CS}^4 m_\phi^3}}\frac{\Omega_{\rm GW}(f_{\rm min})}{\Omega_{\rm GW}(f_{\rm GW})} \frac{f_{\rm GW}^3}{f_{\rm min}^3}\right)\,.
\end{equation}
For the case $b = 0^\circ$, $l = 0^\circ$ we obtain a lower bound on the axion mass that can be detected with this method,
\begin{equation}
    \label{eq:massell}
    m_\phi \gtrsim 10^{-13}{\rm\,eV}\,\left(\frac{\ell_{\rm CS}}{10^8{\rm\,km}}\right)^{\!-4/3}\!\!,
\end{equation}
corresponding to GW frequencies $f_a \gtrsim 10\,$Hz. This range overlaps with the target of the aLIGO/VIRGO detector~\cite{Abbott:2016xvh}, which we here consider as the primary instrument at which the effects we search can be looked for. Note that Eq.~\eqref{eq:massell} predicts $\ell_{\rm CS} \propto m_\phi^{-3/4}$ at the edge of the bound, which is the same slope which is recovered numerically in Fig.~\ref{fig:figurebound} using the full numerical expressions for the modelling.\footnote{In the recent preprint in Ref.~\cite{Tsutsui:2022zos}, appeared during the completion of this work, the author provided a new mechanism to constraint the $(m_\phi,l_{\rm{CS}})$ parameter space, with the results partially overlapping with the white region in Fig.~\ref{fig:figurebound}.} The upper bound labelled ''Frame-dragging'' and marked by the green hashed region corresponds to the results in Eq.~\eqref{eq:alihaimoud} from Refs.~\cite{AliHaimoud:2011fw, AliHaimoud:2011fw}, the solid red line labelled ``LIGO + VIRGO'' corresponds to the bound obtained from the resonant enhancement of the GW from binary systems using the LIGO/VIRGO O1+O2 data~\cite{Jung:2020aem}, and the solid blue line labelled ``Secondary GWs'' is the bound from the null result in the search for secondary GWs associated with the coalescence of compact object binaries~\cite{Tsutsui:2022zos}. Finally, the plot features a sharp transition with a hatched region for $m_\phi \gtrsim 1.15 \times 10^{-11}\,$eV, labelled ``SGWB'' and corresponding to the frequency $f_{\rm max} \approx 1.4\,$kHz at which binary NSs are supposed to cease contributing to the SGWB. As we do not consider further contributions to the SGWB above the frequency $f_{\rm max}$, we do not speculate on the shape of the bounds, for which the minimum value of the CS coupling that can be probed is $\ell_{\rm CS} \approx 2\times 10^6\,$km for a signal coming from the GC.

\begin{figure}
	\centering
	\includegraphics[width=0.7\textwidth]{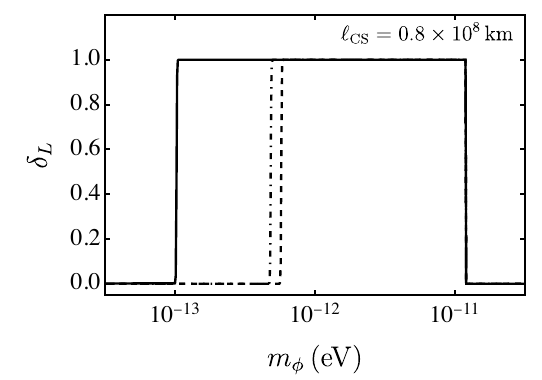}
	\caption{The quantity $\delta_L$ defined in Eq.~\eqref{def_deltaL} as a function of the axion mass $m_\phi$ and for $\ell_{\rm CS} = 0.8\times 10^8\,$km, for different directions of the signal: GC (solid line), Galactic anticenter (dot-dashed line), and perpendicular to the Galactic plane (dashed line).}
	\label{fig:figuredetail}
\end{figure}

The sharp transition of $\delta_L$ between the regions of interest is shown in Fig.~\ref{fig:figuredetail}, in which the CS coupling is fixed to $\ell_{\rm CS} = 0.8\times 10^8\,$km and the same color scheme as in Fig.~\ref{fig:figurebound} is used for the lines. More quantitatively, the transition would occur over a region of mass $\Delta m_\phi \sim \left({\rm d}\ln \Gamma_{\rm GW}/{\rm d}m_\phi\right)^{-1} \sim m_\phi/300$, where in the last expression we have used Eq.~\eqref{eq:approx}. This corresponds to the fractional change in the frequency $\Delta f/f \sim 0.3\%$.

Even for a change in the SGWB of several orders of magnitude, the relation between $\ell_{\rm CS}$ and $m_\phi$ in Eq.~\eqref{eq:approx} would not be altered significantly. On one hand, this leads to considering the findings in Fig.~\ref{fig:figurebound} as a solid result even in the case where exotic components that have not been included in the analysis would exist. However, this also leads to the consideration that the most significant contribution from new physics to the results obtained comes from extending the region to higher masses and frequencies. This feature is present in the SGWB resulting from cosmic strings within superstring theory and from the breaking of a local U(1) symmetry, which are expected to radiate predominantly in GWs~\cite{Olum:1999sg, Moore:2001px}.

For an exhaustive list of exotic sources in astrophysics and cosmology that would lead to detectable signals for frequencies above the kHz to the GHz see Ref.~\cite{Aggarwal:2020olq}.

\subsection{Effects on neutrino oscillations}

We now assess the effects of CS gravity on neutrino flavor oscillations. Given the emission probability $P_{\sigma}(0)$ for the flavor $\sigma$ produced by a source at a distance $D$ from Earth, the probability to detect the flavor $\lambda$ is~\cite{Dvornikov:2021sac}
\begin{eqnarray}
    \label{eq:probability}
    P_{\lambda}(D) &=& \sum_\sigma P_{\sigma}(0)\bigg[\sum_i|U_{\lambda i}|^2|U_{\sigma i}|^2\\ 
    && + 2{\rm Re}\sum_{i>j}U_{\lambda i}U^*_{\lambda j}U^*_{\sigma i}U_{\sigma j}e^{-i D \frac{\Delta m_{ji}^2}{2E}}\,e^{-D/L_{\rm coh}}\bigg]\,,\nonumber
\end{eqnarray}
where the mixing matrix $U$ is defined in Eq.~\eqref{eq:mixingmatrix} and the indices $i$, $j$ label the mass eigenstates. A similar expression holds for the probability in the absence of a CS term when replacing $L_{\rm coh} \to L_{\rm coh,0}$ in Eq.~\eqref{eq:probability}. The expression above can also be understood in terms of the evolution of the neutrino matrix density in Eq.~\eqref{evolution1}.\footnote{In the illustrative case of two active neutrino flavors, setting the initial condition $\rho_{\mathrm{I}}(0)_{11}=P_{e}(0)$ and $\rho_{\mathrm{I}}(0)_{22}=P_{x}(0)=1-P_{e}(0)$, where $P_{e}(0)$ and $P_{x}(0)$ are the initial probabilities of emitting $\nu_{e}$ and $\nu_{x}$, and $\rho_{\mathrm{I}}(0)_{12}=\rho_{\mathrm{I}}(0)_{12}=0$ (no correlations between the initial fluxes of different flavors), the density matrix at a later time $t$ is~\cite{Dvornikov:2019jkc}
\begin{eqnarray}
  \langle \rho_{\mathrm{I}} \rangle_{11} & = & P_e (0) +\frac{1}{2}\sin^{2}(2\theta)(P_{x}(0)-P_{e}(0)) \left[ 1-e^{-\Gamma t} \right], \nonumber  \\
  \langle \rho_{\mathrm{I}} \rangle _{22} & = & P_{x}(0) -\frac{1}{2}\sin^{2}(2\theta)(P_{x}(0)-P_{e}(0)) \left[ 1-e^{-\Gamma t} \right], \nonumber  \\
  \langle \rho_{\mathrm{I}} \rangle _{12} & = & \left\langle \rho_{\mathrm{I}} \right\rangle _{12}= \frac{1}{4}\sin(4\theta)(P_{x}(0)-P_{e}(0))  \left[1-e^{-\Gamma t}  \right],\nonumber
\end{eqnarray}
where $\Gamma = 3 (\Delta m^2)^2 \tau \langle h^{2} \rangle /(256 E^2)$ describes the relaxation of the density matrix over the correlation time $\tau$, $\theta$ is the mixing angle, and $\Delta m^2$ the mass splitting squared.}

As discussed in the previous section, the coherence length computed with a non-zero CS coupling switches from the value $L_{\rm coh} = L_{\rm coh,0}$ obtained in GR to $L_{\rm coh} \ll L_{\rm coh,0}$ when the effects of the CS coupling become important. This corresponds to the change of the quantity $\delta_L$ in Eq.~\eqref{def_deltaL} from $\delta_L = 0$ when the CS coupling does not affect the propagation to $\delta_L \approx 1$ at the onset of the enhancement effect, see Fig.~\ref{fig:figurebound}. The expression in Eq.~\eqref{eq:probability} is composed of an irreducible term plus an oscillatory part, that describes the oscillations of the neutrino flavors with the oscillation length $2E/\Delta m_{ji}^2$. The effect of the coherence length is that of modulating the amplitude of the flavor oscillations, so that when the source is placed at a distance larger than the coherence length the flavor conversion is suppressed. This occurs for $D \gg L_{\rm coh}$ or, given that $L_{\rm coh} \propto E_\nu^2$, below a certain neutrino energy threshold
\begin{equation}
     \label{eq:threshold}
    E_{\nu}^{\rm thr, 0} \approx \left(\frac{9H_0^2D}{1024\pi^3} (\Delta m_{13}^2)^2\int {\rm d}f\,\frac{\Omega_{\rm GW}(f)}{f^4}\right)^{1/2}\,,
\end{equation}
where we have used Eq.~\eqref{eqevolutionrho} and where ``0'' stands for the absence of the CS coupling. A non-zero CS coupling modifies this energy threshold so that for $L_{\rm coh} < L_{\rm coh,0}$ results in $E_\nu^{\rm thr} > E_\nu^{\rm thr, 0}$. This latter condition is satisfied for the same inequality as in Eq.~\eqref{eq:approx}.

\begin{figure}
    \centering
    \includegraphics[width=0.48\textwidth]{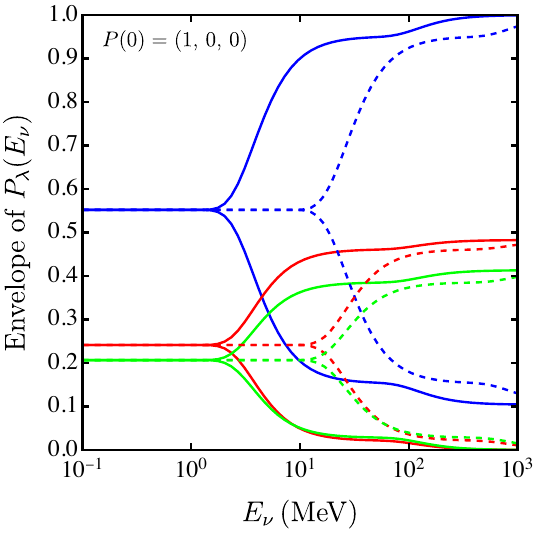}
    \includegraphics[width=0.48\textwidth]{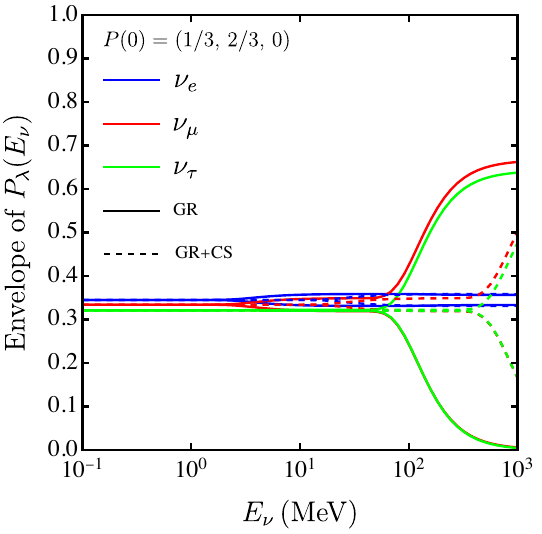}
    \caption{The envelope of the probability $P_\lambda(D)$ in Eq.~\eqref{eq:probability} to observe a neutrino of flavor $\lambda$ at Earth from a source at a distance $D = 1\,$kpc as a function of the neutrino energy $E$, for the initial flavor source $P_\sigma(0) = (1,0,0)$ (left panel) and $P_\sigma(0) = (1/3,2/3,0)$ (right panel). The colors code the neutrino flavors detected: blue (electron), red (muon), green (tau), for the case of GR only (solid line) and for GR with the addition of the CS coupling (dashed line).}
    \label{fig:figureprob}
\end{figure}

In Fig.~\ref{fig:figureprob} we show the results from computing the envelope of the probability $P_\lambda(D)$ to observe a neutrino of flavor $\lambda$ at Earth, obtained by replacing the factor $\exp(-i D\Delta m_{ji}^2/(2E))$ in Eq.~\eqref{eq:probability} with $\pm 1$, for the observed neutrino flavors coded as electron (blue), muon (red), tau (green). We have fixed the neutrino source at $D = 1\,$kpc and we have considered the neutrino energy range $E_\nu = (10^{-1}, 10^3)\,$MeV. We have shown the amplitude of the oscillations in the probability as a function of energy, described by the second term in Eq.~\eqref{eq:probability}. To detect the effect, it is required a neutrino source with a known theoretical initial distribution. Therefore, it is possible to consider as sources the well studied stellar and supernovae~\cite{Janka:2017vlw,Mirizzi:2015eza} framework and the NS merging~\cite{Cusinato:2021zin,Walk:2019miz}. As an example, we consider neutrinos emitted during a SN explosion with two different burst composition: i) a $\nu_e$ burst such that $P_{\nu_e}(0) = 1$ and the other two flavors are initially absent (left panel), and ii) a scenario of source emission in which both $\nu_e$ and $\nu_{\mu}$ are emitted with the composition $(P_{\nu_e}, P_{\nu_\mu},P_{\nu_\tau})= (1/3, 2/3, 0)$ (right panel). The suppression of the oscillation pattern towards lower neutrino energies is due to the effect of the term $\exp(-D/L_{\rm coh})$, due to $L_{\rm coh}\propto E_\nu^2$. The effects of the CS term (dashed lines) lead to an enhancement of the suppression due to the coherence length, particularly at lower energies, with respect to the GR only case (solid line). This corresponds to an increase in the neutrino energy threshold below which the oscillations are suppressed, which for both scenarios increases by a factor $\mathcal{O}(10)$ from $E_\nu^{\rm thr, 0} \approx 2\,$MeV to $E_\nu^{\rm thr} \approx 20 \,$MeV. In these examples, oscillations would not be observed within the neutrino energy range $2-20\,$MeV even if they are expected in the GR scenario. To compute these effects, we have considered $m_\phi = 10^{-12}\,$eV, $\ell_{\rm CS} = 1.6\times 10^{7}\,$km and a direction of the signal coming from the GC. As shown in Fig.~\ref{fig:figurebound}, these values of $(m_\phi, \ell_{\rm CS})$ lie at the boundary between $\delta_L = 0$ and $\delta_L = 1$ where the CS coupling plays a role. Although we do not show the oscillation pattern in the figure, we have checked that for all energies the sum of probabilities is one.

To better understand the energy range and the distance of the source for which the detection is possible, we demand that the oscillation length of the neutrino packet be larger than the spatial resolution of the detector $\sigma_x$ so that it is possible to pin down the conversion in the neutrino flavors within a detection of the signal. At the same time, the coherence length $L_{\rm coh, 0}$ in the absence of a CS coupling has to be larger than the distance $D$ from the source, while it has to be inferior than $D$ in the presence of the CS coupling. Using Eqs.~\eqref{eq:decayrate0} and~\eqref{decayrateCS}, we obtain:
\begin{eqnarray}
	L_{\rm osc}& \gtrsim & \sigma_x \,\ , \\
	D& \lesssim &\frac{256 \pi^3}{3} \frac{L_{\rm osc}^2}{H_0^2} \frac{f_{\rm min}^3}{\Omega_{\rm GW}(f_{\rm min})} \,\ , \\
	D& \gtrsim &\frac{256 \pi^3}{9\sqrt{2\pi\alpha_2}} \frac{L_{\rm osc}^2}{H_0^2}\frac{1}{\ell_{\rm CS}^2 m_\phi^{3/2} e^{\ell_{\rm CS}^4 m_\phi^3\alpha_1}}\frac{f_{\rm GW}^3}{\Omega_{\rm GW}(f_{\rm GW})} \,\ .
\end{eqnarray}
Combining these requirements gives
\begin{equation}
	\label{eq:thresholdmin}
	 \max\left\{\sigma_x, \left(\frac{3H_0^2\,D}{256\pi^3}\frac{\Omega_{\rm GW}(f_{\min})}{f_{\min}^3}\right)^{1/2}\right\}\lesssim L_{\rm osc} \lesssim \left(\frac{9\sqrt{2\pi\alpha_2}}{256 \pi^3} \frac{H_0^2}{L_{\rm osc}^2} \ell_{\rm CS}^2 m_\phi^{3/2} e^{\ell_{\rm CS}^4 m_\phi^3\alpha_1}\frac{\Omega_{\rm GW}(f_{\rm GW})}{f_{\rm GW}^3}\right)^{1/2} \,.
\end{equation}
The second term of the left member is larger than $\sigma_x$ if the source is at the minimum distance $D > 8{\rm\,pc}\,(\sigma_x/{\rm km})^2$, allowing us to consider all extra-solar neutrino sources that lie within our Milky Way. Therefore, being in the latter hypothesis, setting $L_{\rm osc} = 2E_\nu/\Delta m_{13}^2$, the energy range where is possible to detect a different behaviour of the neutrinos respect to GR is:
\begin{equation}
	\label{eq:thresholdmin1}
	  \left(\frac{3H_0^2\,D}{256\pi^3}\frac{\Omega_{\rm GW}(f_{\min})}{f_{\min}^3}\frac{\Delta m_{13}^2}{2}\right)^{1/2}\lesssim E_\nu \lesssim \left(\frac{9\sqrt{2\pi\alpha_2}}{256 \pi^3} \frac{H_0^2}{L_{\rm osc}^2} \ell_{\rm CS}^2 m_\phi^{3/2} e^{\ell_{\rm CS}^4 m_\phi^3\alpha_1}\frac{\Omega_{\rm GW}(f_{\rm GW})}{f_{\rm GW}^3}\frac{\Delta m_{13}^2}{2}\right)^{1/2} \,.
\end{equation}

The energy range considered in the examples of Fig.~\ref{fig:figureprob} is well within the capabilities underlines for the Super-Kamiokande and JUNO experiments, designed to detect supernovae neutrinos down to energies below the MeV~\cite{Martellini:2019era, Super-Kamiokande:2007zsl} with a nominal positron vertex resolution of $\mathcal{O}(1)\,$m for Super-Kamiokande~\cite{Super-Kamiokande:2019xnm} and $\sigma_x\sim\mathcal{O}(10)~\rm{cm}$ for JUNO~\cite{Liu:2018fpq}. These resolution values allow to consider always correct the assumption on the neutrino source distance, $D > 8{\rm\,pc}\,(\sigma_x/{\rm km})^2$. Note, that the JUNO apparatus consisting of a $2\times 10^4\,$t liquid scintillator detector~\cite{JUNO:2015zny}, has the potential to lower the detection threshold to $\sim 20$\,keV~\cite{JUNO_MM_trigger}, therefore allowing to analyze the whole neutrino spectrum expected from SNe. In particular, the effects of the change in the neutrino flavor oscillation patterns described in Fig.~\ref{fig:figureprob} are accessible by the two detectors.

The effects of the cutoff in the coherence length would be evident in the low energy spectrum, since, in the absence of the CS effect, for sufficiently low energy smaller than $E_\nu^{\rm thr}$ it would be possible to observe the neutrino oscillation suppression caused by the presence of the SGWB. Also, the latter effect would be possible to observe: considering for example the JUNO experiment, the energy resolution is $\sim 3\% \sqrt{{\rm MeV}/E}$~\cite{JUNO:2015zny, Steiger:2022vrk} so that the neutrino search is more sensitive towards the higher end of the spectrum, within the threshold $E_\nu \sim 50\,$MeV.

The presence of the CS term would lead to the suppression of the flavor oscillations in the neutrino spectrum range $E_\nu \gtrsim E_\nu^{\rm thr}$. This would happen whenever the combination $(m_\phi, \ell_{\rm CS})$ falls within the region highlighted in Fig.~\ref{fig:figurebound} which, for a signal coming from the GC, corresponds approximately to the constraint in Eq.~\eqref{eq:massell}. The treatment here exposed is also valid for chirping GWs emitted from merging binaries. In that events, the characteristics of the (non)-observation of neutrino flavor oscillations at Earth would constraint or hint at the properties of the parameter space of the CS gravity $(m_\phi,\ell_{\rm CS})$.

\section{Conclusions}
\label{Conclusions}

We have studied the effects of a stochastic background of GWs on the propagation of neutrinos for a gravitational model described by General Relativity plus an additional axion-gravity Chern-Simons coupling term. In Chern-Simons gravity, the additional axion-gravity coupling stimulates a GW burst at a frequency related to the mass of the axion field making up the dark matter halo. Although the gravitational perturbation is typically very small, a resonance peak in the frequency spectrum may arise provided that the GW frequency matches the axion Compton frequency~\cite{Jung:2020aem}. Using Eq.~\eqref{evolution1}, we have calculated the effect of the enhanced GW on the neutrino flavor oscillations, assuming a random distribution of the sources yielding the stochastic GW spectrum. We have numerically solved Eq.~\eqref{evolution} showing how the neutrino flavor oscillations are suppressed in some energy band in the presence of cold and light axions.

The combinations of the CS coupling $\ell_{\rm CS}$ and the axion mass $m_\phi$ for which the decoherence of an otherwise coherent neutrino wave packet is observed at Earth is described by the parameter $\delta_L = 1$ in Eq.~\eqref{def_deltaL} as shown in Fig.~\ref{fig:figurebound}. For a neutrino burst from a supernova at a distance $D$, the energy range for which this phenomenon is detectable in neutrino detectors such as JUNO and Super-Kamiokande is found to be $E_\nu \gtrsim E_\nu^{\rm thr}$ with $E_\nu^{\rm thr} \approx \mathcal{O}(10)\,$MeV, showing that the phenomenon is potentially observable in future SNe explosions. A difference in the neutrino energy threshold below which flavor oscillations are suppressed with respect to what is expected from the SGWB of known astrophysical objects could hint at the presence of CS gravity, as discussed in relation with Fig.~\ref{fig:figureprob}. In conclusion, the pattern of flavor neutrino oscillations can test fundamental particle physics and gravitational theories. Moreover, the precise measurements of the neutrino flux will be crucial in future experiments and could reveal the existence of the axion or constrain their parameter space.

\acknowledgments
The work of G.L.\ and L.M.\ is supported by the Italian Istituto Nazionale di Fisica Nucleare (INFN) through the ``QGSKY'' project and by Ministero dell'Istruzione, Universit\`a e Ricerca (MIUR). The computational work has been executed on the IT resources of the ReCaS-Bari data center, which have been made available by two projects financed by the MIUR (Italian Ministry for Education, University and Research) in the ``PON Ricerca e Competitivit\`a 2007-2013'' Program: ReCaS (Azione I - Interventi di rafforzamento strutturale, PONa3\_00052, Avviso 254/Ric) and PRISMA (Asse II - Sostegno all'innovazione, PON04a2A).

\appendix

\section{Neutrino in Chern-Simons geometry}
\label{appendix}

The presence of an additional CS term modifying GR directly affects the propagation of the neutrinos through the covariant derivatives appearing in the action
\begin{equation}
    \label{eq:DiracAction}
    S_D = \frac{1}{2}\int {\rm d}^4 x\,\sqrt{-g}\left(i\bar\nu\gamma^\mu {e_\mu}^aD_a\nu + {\rm c.c.} + m_i\bar\nu\nu\right)\,,
\end{equation}
where the connections appearing in the covariant derivative are parametrized as~\cite{Perez:2005pm}
\begin{equation}
    D_a\nu = \partial_a\nu + \frac{1}{4}\left({\omega_a}^{\mu\sigma} + {C_a}^{\mu\sigma}\right)\gamma_\mu\gamma_\sigma\nu\,.
\end{equation}
Here, ${\omega_a}^{\mu\sigma}$ is a symmetric and torsion-free connection that depends only on the tetrad ${e_a}^\mu$, while ${C_a}^{\mu\sigma}$ is the contorsion tensor that induces a gravitational coupling between the neutrino and the axion over a conformally flat spacetime as~\cite{Alexander:2008wi}
\begin{equation}
    \label{eq:neutrinolagrangian}
    S_D \supset -\frac{\ell_{\rm CS}^2}{16}\int {\rm d}^4 x\,\sqrt{-g}\left(\bar\nu\gamma_5\gamma^b\nu\right)\left[2(\partial^a\theta) R_{ab} \!-\! (\partial_b\theta) R\right]\,,
\end{equation}
where $\theta = \kappa \phi$. The term in Eq.~\eqref{eq:DiracAction} adds up to the action in Eq.~\eqref{eq:action} and describes an effective axion-neutrino coupling which could modify the propagation of neutrinos~\cite{Huang:2018cwo}. In practice, the neutrino-axion coupling depends on the Ricci tensor and scalar, which are negligible in the interstellar space and would suppress the effect. In more detail, with the approximation $R \sim \kappa^2\rho_{\rm DM}$, the interaction term reads $\mathcal{L} \sim g_{\phi\nu}(\partial_a\phi)\bar\nu\gamma_5\gamma^a\nu$ with the coupling $g_{\phi\nu} = \ell_{\rm CS}^2\kappa^3\rho_{\rm DM} \lesssim 10^{-53}{\rm\,eV^{-1}}$ that is vanishing compared to the present detection techniques~\cite{Huang:2018cwo}. In practice, the neutrino-axion coupling depends on the Ricci tensor and scalar, which are negligible in the interstellar space and would suppress the effect. The results could be different in a cosmological setup, where the term could lead to the production of a thermal axion population from neutrino annihilation.

The Lagrangian term in Eq.~\eqref{eq:neutrinolagrangian} also induces an effective Majorana neutrino mass of the order $m_\nu \sim \ell_{\rm CS}^2 \kappa R \sqrt{\rho_{\rm DM}} \sim \kappa^3\ell_{\rm CS}^2\rho_{\rm DM}^{3/2}$, where in the last step we have again adopted the approximation $R \sim \kappa^2\rho_{\rm DM}$. The neutrino mass term generated from this coupling is $m_\nu \lesssim 10^{-55}\,$eV for the DM halo of the MW, justifying neglecting any term that induces a mixing between axion field, neutrino field and the curvature $R$ in the treatment.

\bibliographystyle{JHEP.bst}
\bibliography{sources.bib}

\end{document}